\documentclass[%
 reprint,
 amsmath,amssymb,
 aps,
]{revtex4-2}

\usepackage{graphicx}
\usepackage{dcolumn}
\usepackage{bm}
\usepackage{hyperref}

\begin{document}

\preprint{APS/123-QED}

\title{Alignment of air showers 
produced by ultra-high energy cosmic rays
at the Pierre Auger Observatory }
\thanks{Alignment in cosmic ray particles}%

\author{C. N. Navia}%
 \email{carlos\_navia@id.uff.br}
 \author{M. N. de Oliveira}
\email{marcel\_oliveira@id.uff.br}
\affiliation{
 Instituto de Física, Universidade Federal Fluminense, 24210-346, Niterói, RJ, Brazil}%


\author{A. A. Nepomuseno}
\affiliation{Departamento de Ciências da Natureza, Universidade Federal Fluminense, 28890-000, Rio das Ostras, RJ, Brazil
 }%


\date{\today}

\begin{abstract}
We show that the energy-weighted angular (zenith, azimuth) distribution of extensive air showers (EAS), produced by Ultra High Energy (UHE) cosmic rays at the Pierre Auger Observatory (PAO), has a thrust axis almost transverse to the interplanetary magnetic field (IMF), with a thrust value $Tp \geq 0.64$  ( where 1.0 means a perfect alignment and 0.5 isotropy). This behavior strongly suggests an effect of the IMF on the charged shower particles, producing additional lateral scattering. 
We discuss the weakening of the Earth's magnetic field during geomagnetic storms (30\% of observational time) when the IMF becomes preponderant, strengthening the alignment.
\end{abstract}

\maketitle


\section{\label{sec:introduction}Introduction}

In the 30s P. Auger and collaborators \cite{auge39}, using two detectors with a distance of many meters between them, noticed that they were triggered simultaneously,
indicating the arrival of particles in coincidence. It is the discovery
of the particle shower. The incidence of a high-energy cosmic ray in
the atmosphere produces an extensive air shower of secondary particles.

Several cascade processes generate an EAS, such as particle production (hadrons), in the successive interactions of a primary cosmic ray with the atmospheric nuclei and the cascade process due to the interactions of the secondary particles (hadronic, leptonic, and electromagnetic cascades).
Several scatter processes spread the air showers laterally over several hundreds of meters, so only a ground-based detector array can record them.

In 1995 an international group of researchers 
under the leadership of Cronin
\citep{auge96} began design studies for a new cosmic ray observatory, the Pierre Auger Project. In 2004 the Pierre Auger Observatory (PAO), using a giant hybrid detector, started observing the ultra-high-energy cosmic rays. 
So far, with several results, such as the chemical composition \cite{aab17}, the primary spectrum determination at ultra-high energies, confirming the flattening of the spectrum near $5\times 10^{18}$ eV, the so-called ``ankle'', and also confirming the steepening of the spectrum at around $5\times 10^{19}$ eV \cite{aab20}. The existence of a dipolar anisotropy on the very-high-energy cosmic ray sources \cite{aab18}, gives valuable information on the origin and evolution of the universe.

On the other hand, the detection of particle production in the interaction of two particle beams (in a collider) is within a solid angle close to $4\pi$ sr. A variable that characterizes the event shape is the "thrust" that determines the grain of sphericity of the event.
 We then suggest a generalization of the conventional thrust definition to study the angular (zenith, azimuth) shape distribution of EAS  in the detection plane (ground level).

 Evaluating 22751 EAS, produced by UHE's cosmic rays, with energies above $0.433$ eV, at the PAO data \cite{auge21}, we find a thrust axis meaning an alignment in the EAS's azimuthal structure. 
We show that this alignment is a consequence of IMF's effect on particle showers.

 We also speculate that support for the alignment maybe can occur
during the periodic high-speed solar wind arrival, weakening the Earth's magnetic field. Under this condition, the IMF becomes more predominant on Earth, strengthening the alignment.

 \section{Pierre Auger Observatory data
 }
\label{PAO_data}

The PAO is located in Malargue, Argentina (35$^{\circ}$S; 69$^{\circ}$W, 1400 m a.s.l.). The goal is to study cosmic rays at the highest energy region above EeV (10$^{18}$ eV). Hybrid Detector's construction provides two independent methods to measure the energy, chemical composition, and arrival direction of UHE cosmic rays by cross-calibration.

The Surface Detector Array detects showers of secondary particles produced by collisions of very high-energy cosmic rays with nuclei in the upper atmosphere \cite{abra10}.  The detector consists of 1660 water-Cherenkov detector stations arranged in a 1500 m spaced triangular grid covering a total area of 3000 km$^2$.

The selection of real events and the rejection of random coincidences are made through a hierarchy of triggers \cite{abra10}, which allows the
detection of extensive air showers (EAS) produced by primary cosmic
rays, with energy above $10^{18}$ eV, for zenith angles between 0 and 80 degrees.

A fluorescence light produced by collisions of very high energy cosmic rays with the nitrogen in the upper atmosphere can be detected on moonless nights by the fluorescence detector composed of 27 telescopes housed at four different locations at the edges of the Surface Detector array.
Fluorescence images allow us to reconstruct the cascade's curve and obtain the mass composition and energy of the primary particle \cite{abra10b}, and make a cross-check with the surface array measurements.

On the other hand,  the scaler mode counts all the particles hitting the surface array's detectors without any trigger system \cite{auge11}. As the geomagnetic rigidity cutoff at Malargue is 9.5 GV, incident primary cosmic rays in the upper atmosphere with energies between 10 GeV to 2 TeV produce 90\% of the counting rate in one unit (water Cherenkov detector) \cite{dass12}. Cosmic rays in this energy range are subject to solar modulation \cite{auge11,dass12}. 

The PAO had several upgrades, including an array of muon detectors (AMIGA) \cite{aab21} and the radio wave detection (AERA) associated with EAS \cite{aab18b}.

\section{The IMF at 1 AU}

The IMF is the solar magnetic field carried through the solar wind, into the interplanetary medium, and beyond. Due to the Sun's rotation, the shape of the IMF close to the ecliptic plane is like a rotating spiral structure (Parker spiral) \cite{hand}.

\begin{figure}
\vspace*{-0.0cm}
\hspace*{-1.0cm}
\centering
\includegraphics[width = 3.5in]{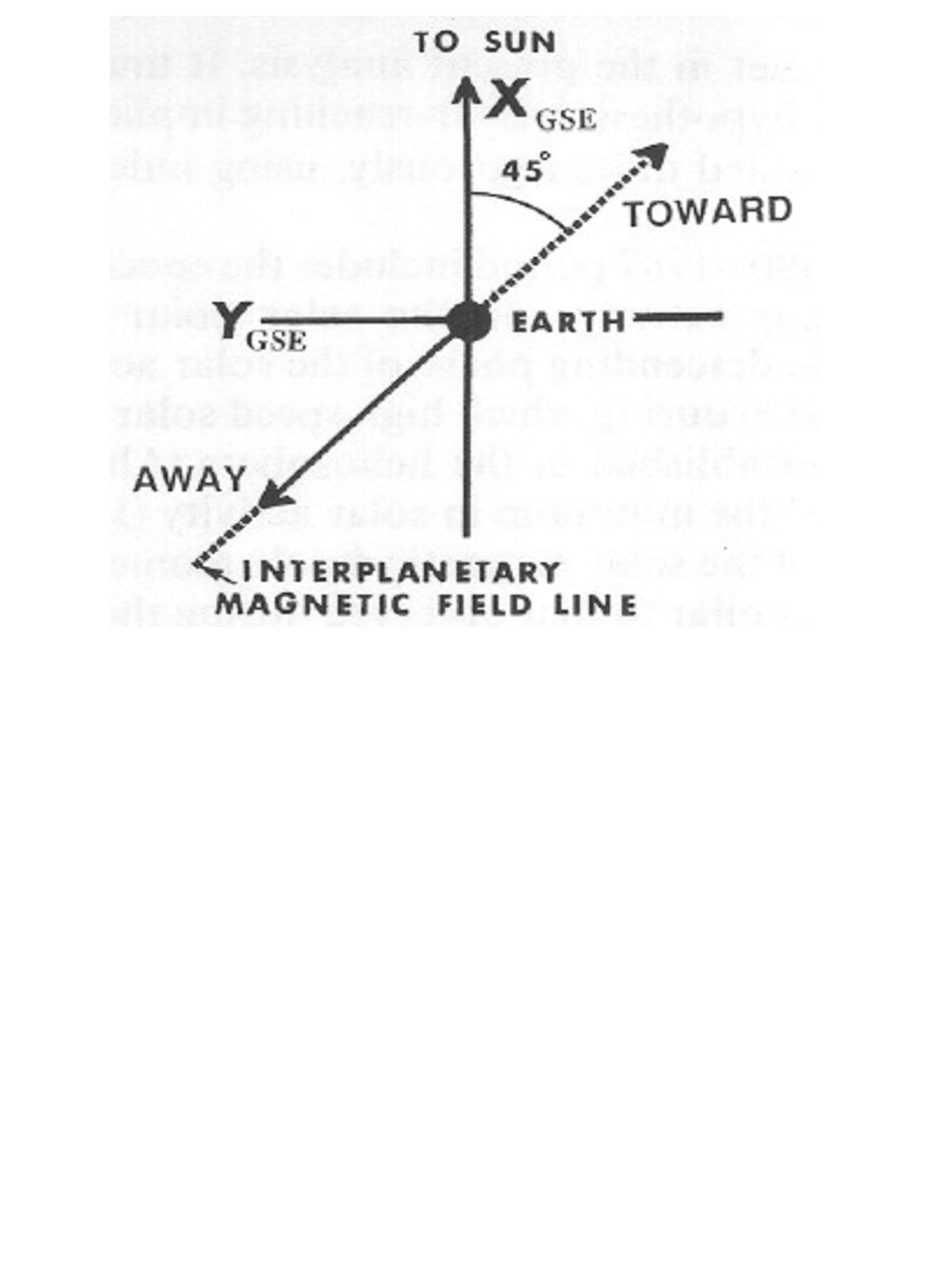}
\vspace*{-6.0cm}
\caption{Geocentric solar ecliptic (GSE) coordinate system. The Sun-Earth line is at 45 degrees in relation to the annual mean orientation of the IMF line.
}
\label{geocentric}
\end{figure} 

 The Sun's magnetic field direction in the northern hemisphere is opposite that of the field in the southern hemisphere. 
 However, this polarity reverse with each solar cycle.

In the 60s Ness and Wilcox \cite{ness65} 
identified the existence of an interplanetary sector boundary, as a surface separating regions of the heliosphere where; the IMF point towards the Sun (negative polarity) or away from it (positive polarity). The surface separating the polarities is called the Heliospheric Current Sheet (HCS). 

Due to an offset between the Sun's rotation axis concerning the magnetic axis, the shape of the Sun's IMF close to the ecliptic plane 
makes the Earth be sometimes above and sometimes below the rotating current sheet. The changes in the polarity of the IMF at 1AU are from two times per Bartels rotation (27 days) around the minimum solar cycle and up to four times around the maximum one.

The IMF total strength ($B_t$) is the result of several magnetic components, 
such as the north-south, west-east, and towards and away from the Sun. The $B_t$ strength increase with the increase in the solar wind, especially with the incidence of an HSS that can trigger geomagnetic storms. So the IMF's rank in the Earth's orbit is from 1 to 37 nT.

On the other hand, the Earth's magnetic field at PAO latitudes is around 24-25 nT in the north direction. However, due to the arrival of an HSS in the magnetosphere triggering a geomagnetic storm, the Earth's magnetic field is weakened, and the IMF field is enhanced \cite{knip93}. This behavior happens during 30\% of the observational time.

Hereafter, we use a system based on the Earth-Sun line, the Geocentric solar ecliptic (GSE) system
\cite{hapg92}.
. This system has its X axis towards the Sun and its Z axis perpendicular to the plane of the Earth's orbit around the Sun, the ecliptic plane  (positive North), Fig.~\ref{geocentric} illustrate the situation. The Earth-Sun line is at 45 degrees from the annual mean direction of the IMF line. Fig.~\ref{geocentric} also indicates the polarity criteria away and toward, adopted in the GSE system.

 \begin{figure}
\vspace*{-0.0cm}
\hspace*{-0.5cm}
\hspace*{-0.0cm}
\centering
\includegraphics[width = 3.5 in]{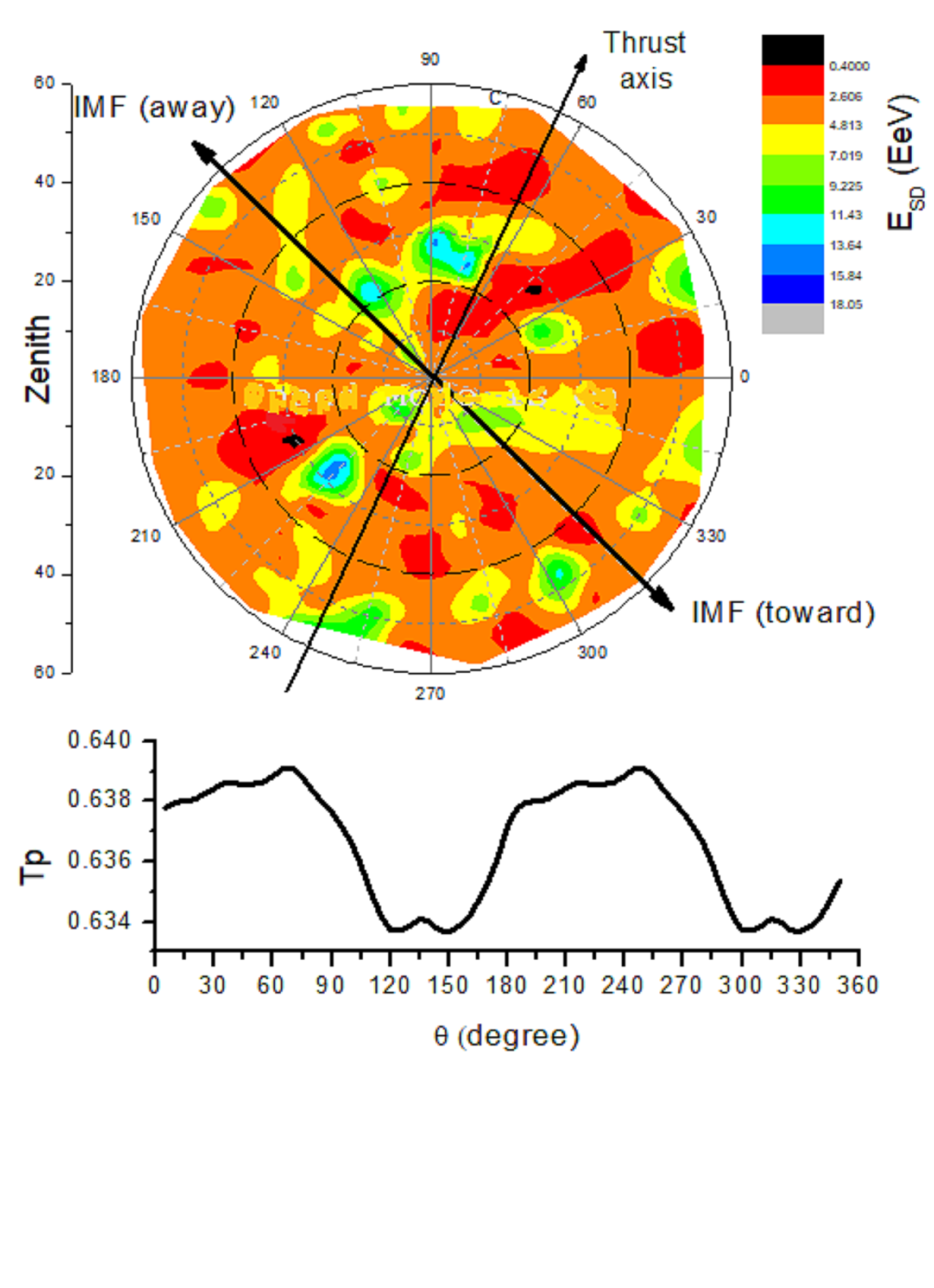}
\vspace*{-2.5cm}
\caption{Top panel: Polar contour plot (zenith and azimuth) weighted according to energy for 22284 EAS in the energy band
$0.43 < E_{SD} < 18$ EeV from PAO's data.
The IMF average annual direction and the thrust axis are indicated for the long arrows. Bottom panel:  Correlation according to Eq.~\ref{thrust}, the $\theta$ values that maximize the function, corresponds to the thrust's axis azimuthal angle.
}
\label{azimuth1}
\end{figure} 

\begin{figure}
\vspace*{-0.0cm}
\hspace*{-0.5cm}
\hspace*{-0.0cm}
\centering
\includegraphics[width = 3.5 in]{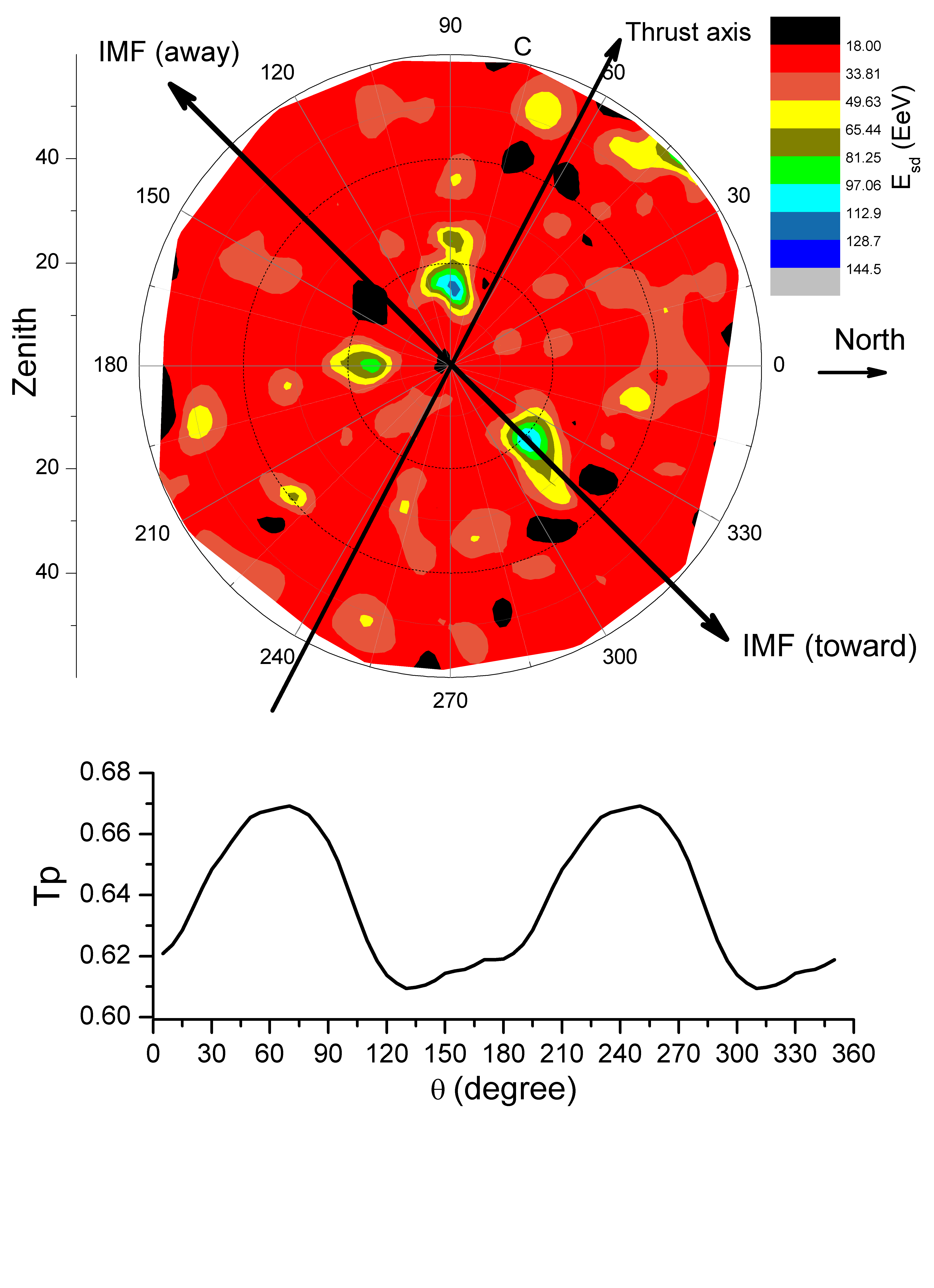}
\vspace*{-2.0cm}
\caption{The same as Fig.~\ref{azimuth1}, but for 465 EAS with energies $E_{SD}>18$ EeV.
}
\label{azimuth2}
\end{figure} 

\begin{figure}
\vspace*{+0.5cm}
\hspace*{-0.5cm}
\hspace*{-0.0cm}
\centering
\includegraphics[width = 3.5 in]{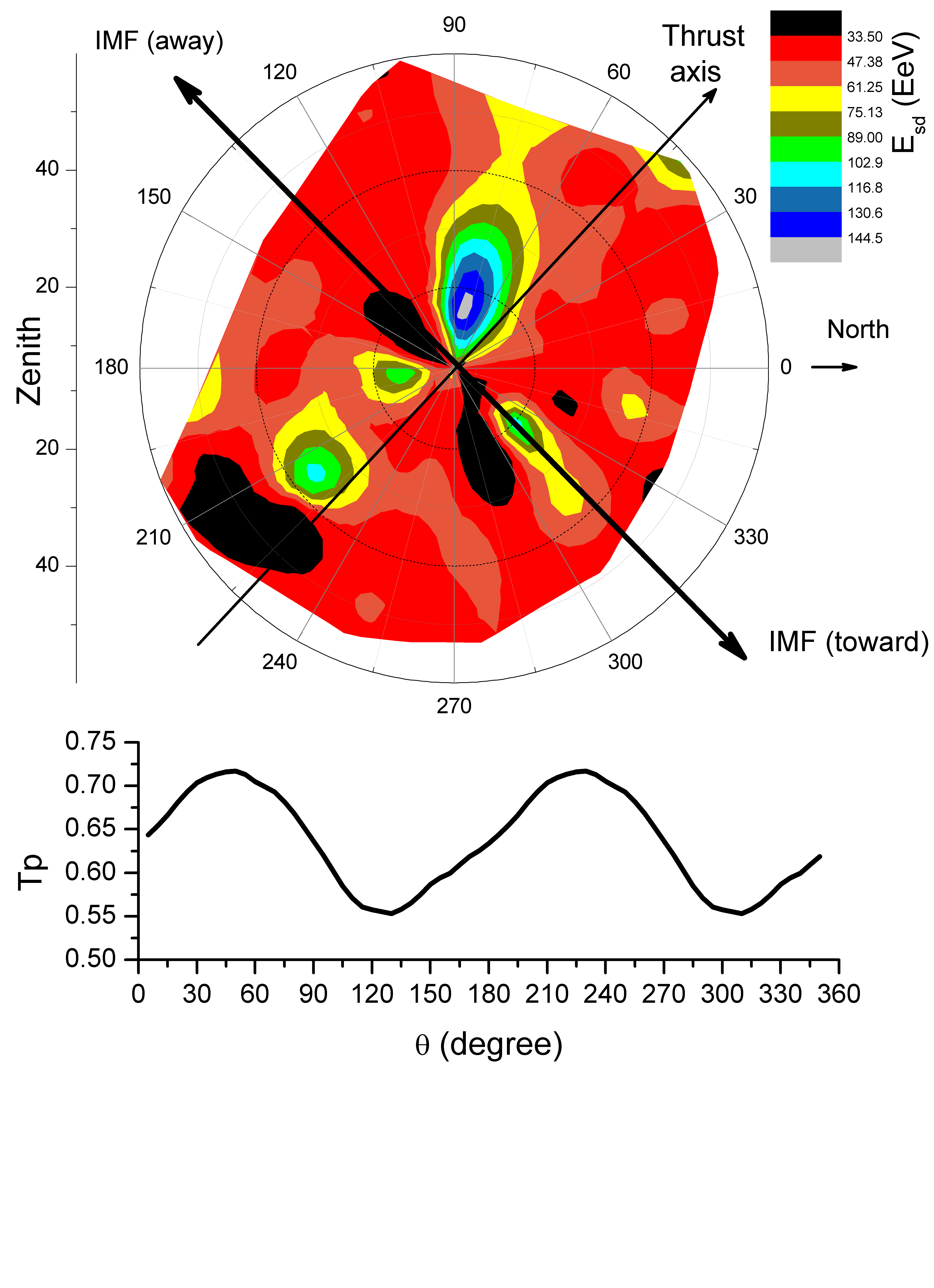}
\vspace*{-2.5cm}
\caption{The same as Fig.~\ref{azimuth1}, but for 85 EAS with energies $E_{SD}>33.5$ EeV.
}
\label{azimuth3}
\end{figure} 

\begin{figure}
\vspace*{+0.0cm}
\hspace*{-1.0cm}
\centering
\includegraphics[width = 3.5in]{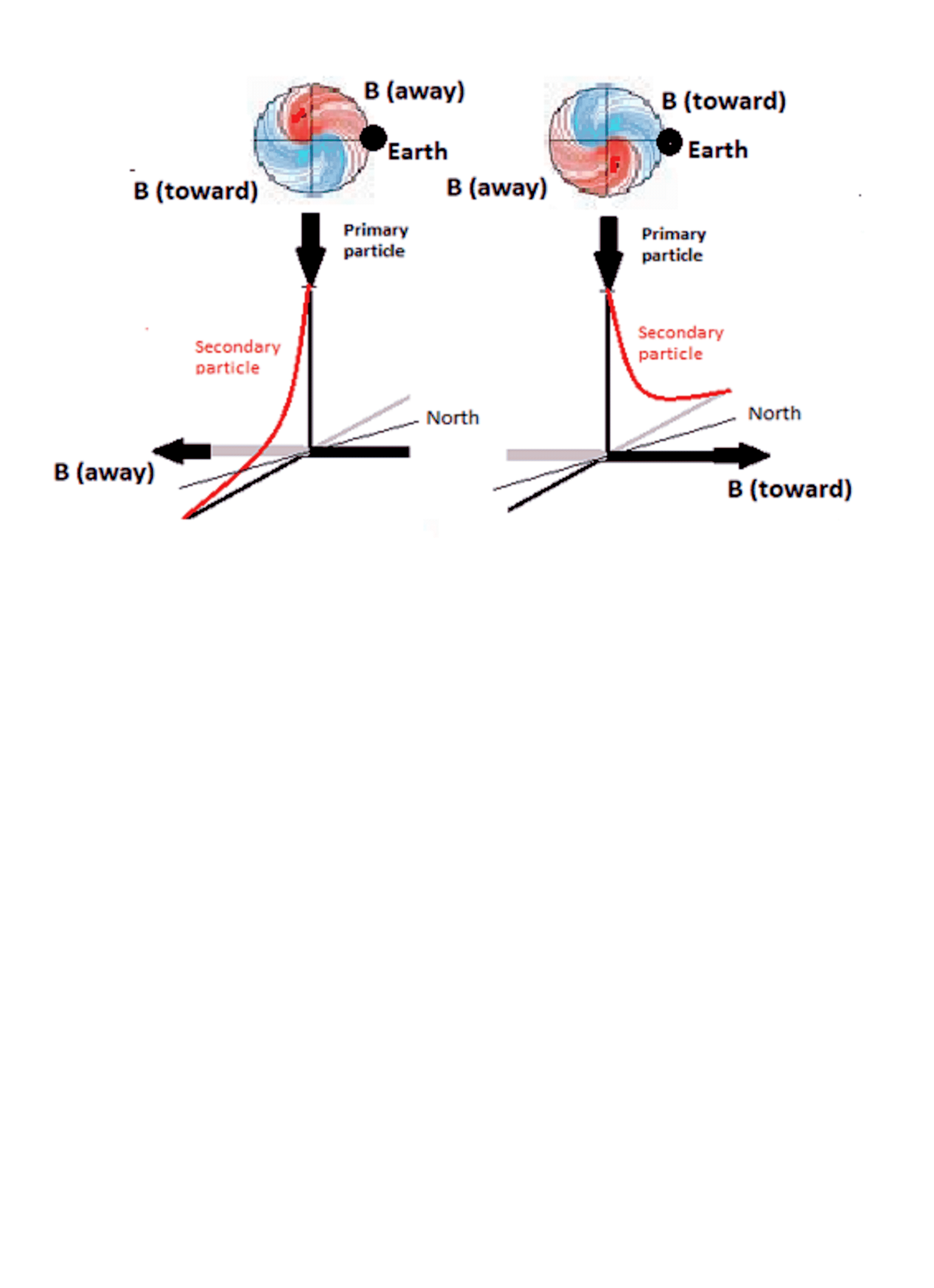}
\vspace*{-7.0cm}
\caption{Diagram of the dispersion of positively charged shower particles under the influence of the IMF. The scatter of particles is in the direction transverse to the IMF.
Changes in IMF polarity; only reverses the scattering side of the shower particles. IMF polarity changes happen when the Earth crosses different magnetic sectors (toward or away from the Sun),
   as shown at the top of the figure.
}
\label{lorentz}
\end{figure} 

\section{Results}

The atmospheric parameter variation, such as the 
barometric pressure and density can influence the air shower's longitudinal development and lateral spread. The event rate of air showers on the PAO's surface detector array shows a $\sim 10\%$ of seasonal modulation and a $\sim 2\%$ due to the diurnal variation \cite{abra10}.

Early papers \cite{cocc54,norm56}, and references therein show that the displacement
of particles from the shower axis by the Earth’s magnetic field is about 2\%, much smaller, in comparison with Coulomb scattering. However, also the data shows a variation in the azimuth,
an ellipticity shape in the shower azimuthal structure.

The PAO'data shows that the distribution of the shower size when considering
the Earth's magnetic field effect is about 2\% higher than the
shower size without considering it \cite{abre11}. In addition, the geomagnetic field in showers mimics a pseudo-dipolar pattern, with small implications 
in the arrival direction distributions of cosmic rays \cite{abre11}. 

Here, firstly we look for some asymmetries in the polar contour plots (zenith and azimuth) of EAS from PAO's surface array, weighted according to the energy associated  ($E_{SD}$), we use the full available data \cite{auge21}. They are 22721 EAS, from cosmic ray primaries with energies ($E_{SD}$)
from $0.433$ EeV up to 144.115 EeV. 

For events with energies below 2.5 EeV, the Auger surface detector trigger is not fully efficient. To overcome this problem, we have analyzed three energy bands:
 (a) $0.433<E_{SD}<18$ EeV, 22284 events, (b) $E_{SD}>18$ EeV, 465 events, and (c) $E_{SD}>33.5$ EeV, 85 events.
The top panel of the figures \ref{azimuth1}, \ref{azimuth2} and \ref{azimuth3} show these polar contour plots. 
From these figures, we can see a tendency towards an azimuthal alignment, almost transverse to the direction of the IMF, especially in the last two figures. 

Secondly, we quantify these alignments using the ``thrust'' variable.
Thrust (T) is a variable used to characterize the collision of high-energy particles at collider physics \cite{bran64,cesa21}. When two high-energy particles collide, they typically produce jets of secondary particles. The thrust variable indicates the event shape. An event with jet particles spherically 
distributed would have $T=0.5$, and if the jet particles are perfectly alignment, have a $T=1.0$.

However, shower particles produced by cosmic ray interactions, are only recorded in one plane (ground level). 
In this case, using the variable that we call planar thrust (Tp) is more appropriate. That requires replacing the moment with the transverse moment. In our case, it substitutes the energy of the EAS $E_{SD}$  by the so-called transverse energy $E_T=E_{SD} \sin(\theta_i)$, where $\theta_i$ is the zenith angle of the EAS, and the Tp is

\begin{equation}
T_p= max \frac{\Sigma_i^N E_T^i|\cos(\phi_i-\theta)|}{|\Sigma_i^N E_T^i|},
\label{thrust}
\end{equation}
where N is the number of showers, $E_T^i $, is the transverse energy of the shower \textit{i}, 
$\theta_i$ and $\phi_i$ are the zenith and azimuth angles of the shower \textit{i}, and
$\theta$ (alone) is the azimuthal angle of the thrust axis that maximizes the sum ratio.
The range of $T_p$ is from $0.5$ (perfectly isotropic)  to $1.0$ (perfectly aligned).

The bottom panels of figures \ref{azimuth1}, \ref{azimuth2}, and \ref{azimuth3} show the Tp variation as a function of $\theta$ according to Eq.~\ref{thrust}. In all cases, the Tp presents two maxima, which define the direction of the thrust axis in the polar contour plot. The thrust values (maxima) are $\sim$0.64 for $0.43<E_{SD}<18$ EeV, 0.67 for $E_{SD}>18.06$ EeV, and 0.72 for $E_{SD}> 33.5$ EeV. That means that the alignment increase as the $E_{SD}$ increase.

In addition, changes in IMF polarity (toward-away) only reverse the scattering side of the shower particles, as shown in Fig.~\ref{lorentz}. IMF polarity changes happen when the Earth crosses different magnetic sectors (toward or away from the Sun),
   as shown at the top of Fig.~\ref{lorentz}.

\section{High speed streams effect}
\label{HSS_efect}

Coronal holes (CH) and coronal mass ejections (CME) are the sources of HSS \cite{rich18,cran02,fuji16}. 
According to NOAA Space Center, around 1454 days per solar cycle (11 years) are under geomagnetic storm conditions due to the arrival of an HSS. Under this condition, the IMF strength on Earth becomes above 
10.0 nT and the Earth's magnetic field becomes weakened.



\begin{figure}
\vspace*{+0.0cm}
\hspace*{-0.0cm}
\centering
\includegraphics[width = 3.5in]{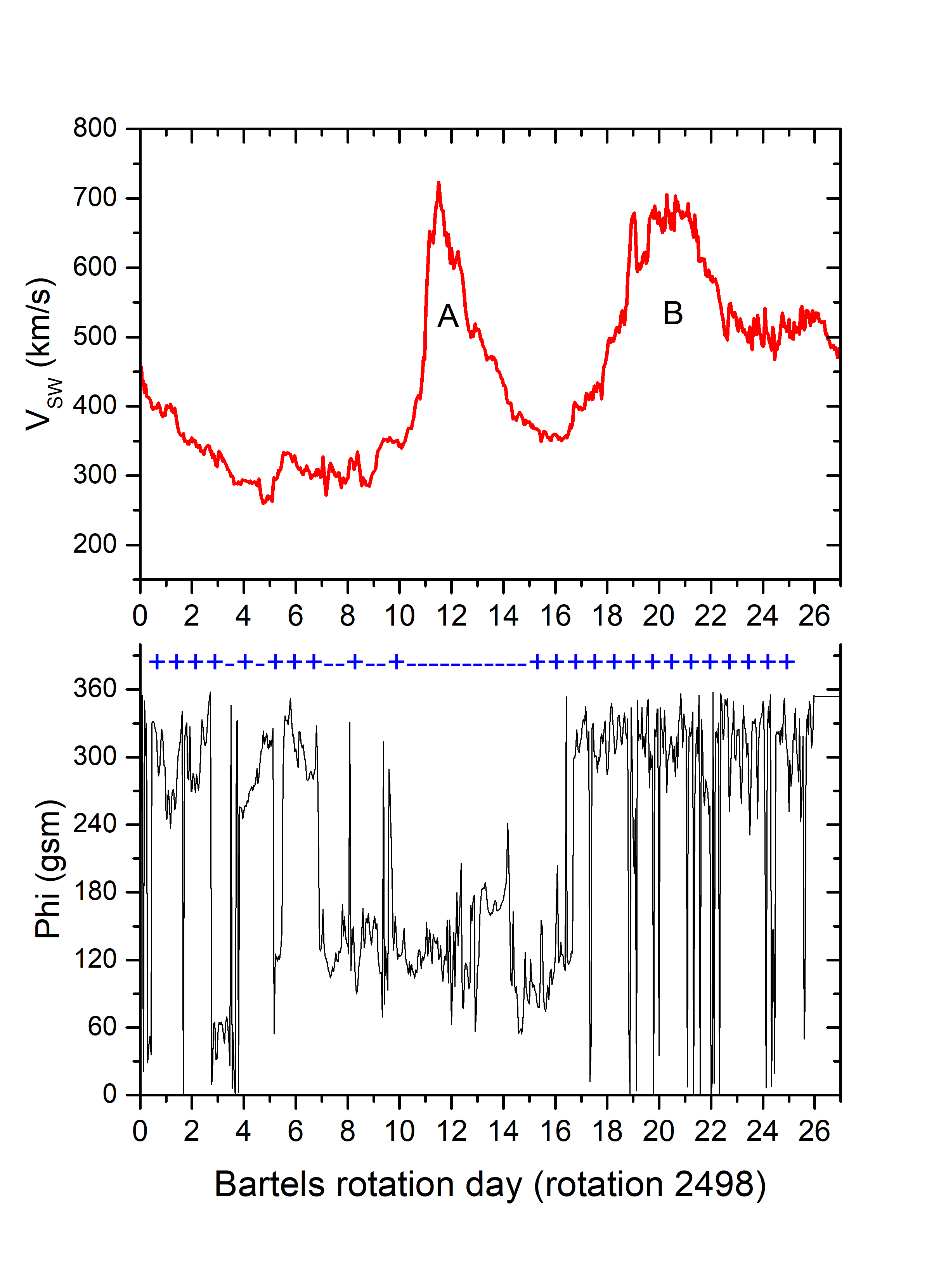}
\vspace*{-1.0cm}
\caption{Top panel: Solar wind speed time profiles, for Bartels 2498 rotation, according to ACE Swepan data, at Lagrange L1 Point. The two HSS reaching speeds above 600 km/s are labeled A and B. Bottom panel: Same as on top, but for the phi magnetic component, according to ACE magnetic data. The polarities, negative (toward Sun) and positive (away Sun), are indicated.
}
\label{HSS}
\end{figure}

In most cases, at Lagrange L1 point, there is one HSS for each IMF's polarity sector, per Bartels rotation.
An example of this behavior is plotted in Fig.~\ref{HSS} (upper panel) showing the solar wind speed time profile for the Bartels 2498 rotation, according to ACE Swepan data \cite{ace}. The two HSS are labeled as A and B. Both reach speeds above 600 km/s.
In addition, we show the corresponding time profile of the polarity of the azimuthal component of the IMF in the lower panel, according to the ACE magnetic data \cite{ace}. In short, each HSS happens in each sector of the IMF polarity.

The shock wave at the front of HSS cracks the magnetosphere,
allowing the injection of protons and electrons from HSS 
into the magnetosphere, forming an equatorial ring current around Earth. The ring current is responsible for a magnetic field directly opposite Earth's magnetic field, weakening it. So the IMF's lines become predominant on Earth. 
The strength of the ring current is measured by the disturbance storm time (Dst) index. Negative values of the Dst index indicate a weakened Earth's magnetic field.
The more negative the Dst index, the weaker the Earth's magnetic field. 

In short, when a magnetic storm is in progress, the Earth's magnetic field gets weakened, and the IMF becomes predominant. Under this condition, the additional lateral scattering due to IMF is more effective. Around 30\% of the observational time is under this condition. However,
due to the complexity, quantification of including HSS on the additional scatter requires a Monte Carlo study.

\section{Conclusions}

From an analysis of the polar contour plots (zenith and azimuth)  of 22751 EAS  weighted according to the energy associated  ($E_{SD}$)  from the PAO's surface detector array data, we showed the existence of a ``thrust'' axis in the polar contour plots almost transverse to the IMF direction.
That means an azimuthal alignment of the showers transverse to the IMF. The data is consistent with a thrust value higher than 0.5 (isotropy) as 0.64 for EAS with $0.43<E_{SD}<18$ EeV and increase as the $E_{SD}$ increases, up to
0.72 for EAS with $E_{SD}>33.5$ EeV.

The fact that alignment is transverse to the IMF strongly suggests a Lorentz magnetic force on the charged shower particles under the influence of the IMF as responsible for an additional lateral scattering. 
Periodic changes in the IMF polarity only reverse the scattering side of the shower particles. 

The first observations of the EAS \cite{cocc54,norm56}, showing an azimuthal structure and a pseudo-dipole pattern in the EAS in the PAO data \cite{abre11}, were attributed to the Earth's magnetic field. According to our results, attributing the IMF for this behavior would be a more plausible explanation.

Also, emulsion chamber experiments at mountain altitudes reported the observation of several alignments on hadron-gamma ray families produced by cosmic ray interaction in the TeV energy region \cite{kope95}.  Our results suggest the IMF can be responsible for these alignments.

The possible implications of the EAS's azimuthal alignment in the arrival direction distribution of cosmic rays require a careful Monte Carlo analysis.

On the other hand, the alignment due to IMF can be more effective with the arrival in the magnetosphere of an HSS, triggering geomagnetic storms. During the progress of a magnetic storm, which happens around 30\%  of observational time, the Earth's magnetic field is weakened, and the IMF is enhanced.

\begin{acknowledgments}
 We express our thanks to the memberships of Pierre Auger Collaboration and the Space Weather Prediction Center (NOAA) for valuable information and open data policy. This work is supported by the Fundação de Amparo à Pesquisa do Estado do Rio de Janeiro (FAPERJ) under Grant E 26/010.101128/2018.
\end{acknowledgments}

\bibliographystyle{unsrt}

\bibliography{bibi}

\begin{thebibliography}{10}

\bibitem{auge39}
Pierre Auger, P.~Ehrenfest, R.~Maze, J.~Daudin, and Robley~A. Fr\'eon.
\newblock Extensive cosmic-ray showers.
\newblock {\em Rev. Mod. Phys.}, 11:288--291, Jul 1939.

\bibitem{auge96}
Pierre~Auger Collaboration.
\newblock Auger project design report.
\newblock
  \url{https://visitantes.auger.org.ar/index.php/auger-project-design-report/},
  1996.

\bibitem{aab17}
The Pierre~Auger Collaboration.
\newblock Inferences on mass composition and tests of hadronic interactions
  from 0.3 to 100 eev using the water-cherenkov detectors of the pierre auger
  observatory.
\newblock {\em Phys. Rev. D}, 96:122003, Dec 2017.

\bibitem{aab20}
The Pierre~Auger Collaboration.
\newblock Measurement of the cosmic-ray energy spectrum above
  $2.5\ifmmode\times\else\texttimes\fi{}{10}^{18}\text{ }\text{ }\mathrm{eV}$
  using the pierre auger observatory.
\newblock {\em Phys. Rev. D}, 102:062005, Sep 2020.

\bibitem{aab18}
The Pierre~Auger Collaboration.
\newblock Large-scale cosmic-ray anisotropies above 4 {EeV} measured by the
  pierre auger observatory.
\newblock {\em The Astrophysical Journal}, 868(1):4, nov 2018.

\bibitem{auge21}
Pierre~Auger Collaboration.
\newblock The pierre auger 2021 open data.
\newblock \url{https://opendata.auger.org/}, 2021.

\bibitem{abra10}
Jiju Abraham, P.~Abreu, Marco Aglietta, E.J. Ahn, D.~Allard, I.~Allekotte,
  J.~Allen, J.~Alvarez-Muñiz, M.~Ambrosio, L.~Anchordoqui, S.~Andringa, Tome
  Anticic, Anna Anzalone, C.~Aramo, Ernesto Arganda, Katsushi Arisaka, Fernando
  Arqueros, Hernán Asorey, Pedro Assis, and R.~Bonino.
\newblock Trigger and aperture of the surface detector array of the pierre
  auger observatory.
\newblock {\em Nuclear Instruments and Methods in Physics Research Section A:
  Accelerators, Spectrometers, Detectors and Associated Equipment},
  613:29–39, 01 2010.

\bibitem{abra10b}
The Pierre~Auger Collaboration.
\newblock Measurement of the depth of maximum of extensive air showers above
  ${10}^{18}\text{ }\text{ }\mathrm{eV}$.
\newblock {\em Phys. Rev. Lett.}, 104:091101, Mar 2010.

\bibitem{auge11}
The Pierre~Auger Collaboration.
\newblock The pierre auger observatory scaler mode for the study of solar
  activity modulation of galactic cosmic rays.
\newblock {\em Journal of Instrumentation}, 6(01):P01003--P01003, jan 2011.

\bibitem{dass12}
S.~Dasso, H.~Asorey, and {For The Pierre Auger Collaboration}.
\newblock The scaler mode in the pierre auger observatory to study heliospheric
  modulation of cosmic rays.
\newblock {\em Advances in Space Research}, 49(11):1563--1569, 2012.
\newblock Advances in theory and observation of solar system dynamics - I.

\bibitem{aab21}
The Pierre~Auger Collabotation.
\newblock Design and implementation of the {AMIGA} embedded system for data
  acquisition.
\newblock {\em Journal of Instrumentation}, 16(07):T07008, jul 2021.

\bibitem{aab18b}
The Pierre~Auger Collaboration.
\newblock Observation of inclined {EeV} air showers with the radio detector of
  the pierre auger observatory.
\newblock {\em Journal of Cosmology and Astroparticle Physics},
  2018(10):026--026, oct 2018.

\bibitem{hand}
Space weather live.
\newblock The interplanetary magnetic field (imf).
\newblock
  \url{https://www.spaceweatherlive.com/en/help/the-interplanetary-magnetic-field-imf.html},
  2022.

\bibitem{ness65}
Norman~F. Ness and John~M. Wilcox.
\newblock Sector structure of the quiet interplanetary magnetic field.
\newblock {\em Science}, 148(3677):1592--1594, 1965.

\bibitem{knip93}
DJ~Knipp, BA~Emery, AD~Richmond, NU~Crooker, MR~Hairston, JA~Cumnock, WF~Denig,
  FJ~Rich, O~de~La~Beaujardiere, JM~Ruohoniemi, et~al.
\newblock Ionospheric convection response to slow, strong variations in a
  northward interplanetary magnetic field: A case study for january 14, 1988.
\newblock {\em Journal of Geophysical Research: Space Physics},
  98(A11):19273--19292, 1993.

\bibitem{hapg92}
M.A. Hapgood.
\newblock Space physics coordinate transformations: A user guide.
\newblock {\em Planetary and Space Science}, 40(5):711--717, 1992.

\bibitem{cocc54}
Giuseppe Cocconi.
\newblock Influence of the earth's magnetic field on the extensive air showers.
\newblock {\em Phys. Rev.}, 93:646--647, Feb 1954.

\bibitem{norm56}
RJ~Norman.
\newblock Influence of the earth's magnetic field on air showers.
\newblock {\em Physical Review}, 101(4):1405, 1956.

\bibitem{abre11}
The Pierre~Auger collaboration.
\newblock The effect of the geomagnetic field on cosmic ray energy estimates
  and large scale anisotropy searches on data from the pierre auger
  observatory.
\newblock {\em Journal of Cosmology and Astroparticle Physics},
  2011(11):022--022, nov 2011.

\bibitem{bran64}
S.~Brandt, Ch. Peyrou, Ryszard Sosnowski, and A.~K. Wr{\'o}blewski.
\newblock The principal axis of jets — an attempt to analyse high-energy
  collisions as two-body processes.
\newblock {\em Physics Letters}, 12:57--61, 1964.

\bibitem{cesa21}
Cari Cesarotti, Matthew Reece, and Matthew~J Strassler.
\newblock The efficacy of event isotropy as an event shape observable.
\newblock {\em Journal of High Energy Physics}, 2021(7):1--36, 2021.

\bibitem{rich18}
Ian~G Richardson.
\newblock Solar wind stream interaction regions throughout the heliosphere.
\newblock {\em Living reviews in solar physics}, 15(1):1--95, 2018.

\bibitem{cran02}
Steven~R Cranmer.
\newblock Coronal holes and the high-speed solar wind.
\newblock {\em Space Science Reviews}, 101(3):229--294, 2002.

\bibitem{fuji16}
K.~Fujiki, M.~Tokumaru, K.~Hayashi, D.~Satonaka, and K.~Hakamada.
\newblock {LONG}-{TERM} {TREND} {OF} {SOLAR} {CORONAL} {HOLE} {DISTRIBUTION}
  {FROM} 1975 {TO} 2014.
\newblock {\em The Astrophysical Journal}, 827(2):L41, aug 2016.

\bibitem{ace}
ACE Real-Time data.
\newblock Ace real-time data.
\newblock \url{https://www.swpc.noaa.gov/products/ace-real-time-solar-wind},
  2022.

\bibitem{kope95}
Kopenkin, Managadze, Rakobolskaya, and Roganova.
\newblock Alignment in gamma -hadron families of cosmic rays.
\newblock {\em Physical review. D, Particles and fields}, 52 5:2766--2774,
  1995.

\end{thebibliography}

\end{document}